\documentclass[%
12pt,
aip,
cha,
amsmath,amssymb,
reprint,
]{revtex4-1}

\usepackage[colorlinks,citecolor=green,linkcolor=blue,bookmarks=false,hypertexnames=true]{hyperref}
\usepackage{lmodern} 

\numberwithin{equation}{section}

\newcommand{\neigh}[1]{\text{neighbor}(#1)}

\usepackage{tikz}
\usetikzlibrary{shapes.geometric,arrows.meta,positioning,chains}
\tikzset{%
  every neuron/.style={
    circle,
    draw,
    minimum size=1cm
  },
  neuron missing/.style={
    draw=none, 
    scale=4,
    text height=0.333cm,
    execute at begin node=\color{black}$\vdots$
  },
}

\begin{document}

\title{Discovering the dynamics of \emph{Sargassum} rafts' centers of mass}

\author{F.J.\ Beron-Vera}
\email{fberon@miami.edu}
\affiliation{Department of Atmospheric Sciences, Rosenstiel School of Marine, Atmospheric \& Earth Science, University of Miami, Miami, Florida, USA.} 

\author{G.\ Bonner}
\email{gbonner@morgridge.org}
\affiliation{Morgridge Institute for Research, University of Wisconsin, Madison, WI, USA.}    

\date{Started: October 15 2024. This version: \today.\vspace{-0.25in}}


\begin{abstract}
    Since 2011, rafts of floating \emph{Sargassum} seaweed have frequently obstructed the coasts of the Intra-Americas Seas. The motion of the rafts is represented by a high-dimensional nonlinear dynamical system. Referred to as the eBOMB model, this builds on the Maxey--Riley equation by incorporating interactions between clumps of \emph{Sargassum} forming a raft and the effects of Earth's rotation. The absence of a predictive law for the rafts' centers of mass suggests a need for machine learning. In this paper, we evaluate and contrast Long Short-Term Memory (LSTM) Recurrent Neural Networks (RNNs) and Sparse Identification of Nonlinear Dynamics (SINDy). In both cases, a physics-inspired closure modeling approach is taken rooted in eBOMB. Specifically, the LSTM model learns a mapping from a collection of eBOMB variables to the difference between raft center-of-mass and ocean velocities. The SINDy model's library of candidate functions is suggested by eBOMB variables and includes windowed velocity terms incorporating far-field effects of the carrying flow. Both LSTM and SINDy models perform most effectively in conditions with tightly bonded clumps, despite declining precision with rising complexity, such as with wind effects and when assessing loosely connected clumps. The LSTM model delivered the best results when designs were straightforward, with fewer neurons and hidden layers. While LSTM model serves as an opaque black-box model lacking interpretability, the SINDy model brings transparency by discerning explicit functional relationships through the function libraries. Integration of the windowed velocity terms enabled effective modeling of nonlocal interactions, particularly in datasets featuring sparsely connected rafts.
\end{abstract}


\maketitle

\begin{quotation}
     Long Short-Term Memory (LSTM) Recurrent Neural Networks (RNNs) are a type of artificial NN designed to model sequential data to capture long-range dependencies. Sparse Identification of Nonlinear Dynamics (SINDy) is a method used for discovering governing equations from data. It identifies a sparse, interpretable model to describe the underlying dynamics of a system by selecting relevant terms from a library of candidate functions. The two machine learning methods are compared in the task of determining a law governing the motion of the center of mass of rafts of \emph{Sargassum} seaweed. Floating on the ocean surface, the movement of \emph{Sargassum} follows a Maxey--Riley formulation where the rafts interact elastically in a nonlinear manner. Yet, the motion of their centers of mass, important in practical applications, remains undetermined, motivating this work.
\end{quotation}

\section{Introduction}

The movement of rafts composed of floating \emph{Sargassum}, a species of brown macroalgae that has recurrently congested the coastlines of the Intra-Americas Seas since approximately 2011 \cite{Wang-etal-19}, is characterized by a high-dimensional nonlinear dynamical system. This system encompasses several modifications to the Maxey--Riley equation employed in fluid mechanics \cite{Maxey-Riley-83, Michaelides-97, Cartwright-etal-10}. First, it considers finite-sized particles situated at the ocean's surface \cite{Beron-etal-19-PoF}. The model, designated as the \emph{BOM model}, has undergone validation through both field \cite{Olascoaga-etal-20, Miron-etal-20-GRL} and laboratory experiments \cite{Miron-etal-20-PoF}, and is reviewed in Beron-Vera \cite{Beron-21-ND}. Furthermore, the Maxey--Riley equation has been augmented to incorporate the elastic interactions among BOM particles, resulting in a system of coupled, nonautonomous first-order ordinary differential equations. These interactions emulate the connectivity between the gas-filled bladders that support the floating \emph{Sargassum} plants via flexible stipes. This adaptation is referred to as the \emph{eBOM} model. The most recent extension of this work, and the primary focus herein, is termed the \emph{eBOMB} model \cite{Bonner-etal-24}. This extension integrates nonlinear elastic interactions, facilitating a representation of the movement of rafts of \emph{Sargassum} plants influenced by physiological changes as they adapt to the combined effects of ocean currents and atmospheric winds. These changes are concurrently addressed by the eBOMB model.

In practical applications such as tracking, prevention, and response, the movement of the rafts' centers of mass is arguably of primary importance. Furthermore, in research focusing on the connectivity of \emph{Sargassum} such as Markov-chain descriptions \cite{Bonner-etal-23}, the trajectories of the rafts' centers of mass are crucial, as they are required to construct the pertinent transition matrices.  However, a first-principles-based law describing this motion does not exist, thus motivating the use of machine learning to derive one. 

In this paper we employ two distinct methodologies to achieve the stated objective, and evaluate their respective performances. The first method utilizes \emph{Long Short-Term Memory} (\emph{LSTM}) \emph{Recurrent Neural Networks} (\emph{RNNs}), an artificial NN variant designed to model sequential data and detect long-range dependencies \cite{Hochreiter-Schmidhuber-97}.  Basically, NNs perform nonlinear regression tasks, employing the gradient descent method (including backpropagation) to adjust network weights and biases while minimizing error.  Such methods may give highly accurate predictions, but suffer from a lack of interpretability, as is the case with many NN architectures. In contrast, the second method, the \emph{Sparse Identification of Nonlinear Dynamics} (\emph{SINDy}) \cite{Brunton-etal-16}, discovers a parsimonious (i.e., sparse) and interpretable (i.e., symbolic) model that captures the intrinsic, generically nonlinear dynamics of a system.  This is done by selecting pertinent terms from a library of candidate functions via the solution of a linear regression problem using sparsity-promoting regularization techniques. The evaluation of the LSTM RNN methodology is conducted by assessing the impact of varying the architecture of the LSTM RNN. For the SINDy methodology, we apply optimal regulation of the number of regressed candidate functions, which include appropriately constructed ``windowed'' functions. These account for the nonlocal nature of the dynamical system controlling the motion of \emph{Sargassum} rafts, where a clump of \emph{Sargassum} within a raft can be influenced not only by neighboring clumps but also by distant ones.  In each case, a physics-informed closure modeling approach is adopted, based on the physics described by the eBOMB model.

The subsequent sections of this paper are organized as follows. Section \ref{sec:ebomb} starts with a review of the eBOM model, followed by an examination of the formulation of a law that governs the motion of the centers of mass of the \emph{Sargassum} rafts. Section \ref{sec:dis} is dedicated to the discovery of such a law, beginning with an explanation of the construction of datasets used for training. Subsection \ref{sec:LSTM} investigates the application of LSTM RNNs for modeling the dynamics of the centers of mass, while Subsection \ref{sec:SINDy} explores this modeling through the use of SINDy. The evaluation of the performance of the developed ``LSTM model'' and ``SINDy model'' is presented in Section \ref{sec:disc}.  The paper concludes with a summary and final remarks in Section \ref{sec:con}. The Supplemental Online Material includes two appendices that offer additional information on machine learning and artificial neural networks, with particular emphasis on LSTM RNNs in Appendix A and SINDy in Appendix B.

\section{The motion of \emph{Sargassum} rafts}\label{sec:ebomb}

According to the eBOMB model \cite{Bonner-etal-24}, a \emph{clump} of \emph{Sargassum} acts as the fundamental unit, affected by both ocean currents and winds. Such a clump is envisioned as a compact sphere with a small diameter, grouping \emph{Sargassum} together. When multiple clumps interact, they form what is called a \emph{raft}. It is hypothesized that the movement of a substantial mass of \emph{Sargassum} can be effectively modeled by the behavior of these distinct clumps. In the context of this study, the entire \emph{Sargassum} formation functions as a raft, potentially comprising several independent networks of clumps. The eBOMB model is structured around three key elements: clump dynamics described by the coupled Maxey--Riley equations, nonlinear spring forces linking clumps together, and a biological model for clump growth and decay, the latter of which is not addressed in this paper.

The physics in the eBOMB model is determined by several parameters. Among these are $\alpha$, a dimensionless indicator of windage; $R$, a dimensionless factor reflecting the spherical clump's interaction with air; and $\tau$, which measures the inertial response, also known as Stokes' time. These parameters depend on the density of the clump compared to water, defined as $\delta \ge 1$, representing buoyancy. The parameter $\tau$ is also influenced by the clump's radius, noted as $a$. Detailed expressions for these parameters are available in the Appendix. Furthermore, three additional parameters influence how clumps interact elastically in a nonlinear manner: $L$, the inherent length of the spring connecting adjacent clumps; $A$, indicating the spring's stiffness amplitude; and $\Delta$, which sets the stiffness cutoff threshold.  

Let $\mathbf x = (x, y)$ denote position on a $\beta$-plane. (While accounting for the full effects of Earth's sphericity is possible \cite{Bonner-etal-24}, we omit this formulation for simplicity. Nonetheless, computations are performed taking these effects into account.) The near-surface ocean velocity and wind at a position $\mathbf x$ and time $t$ are denoted by $\mathbf v(\mathbf x, t)$ and $\mathbf w(\mathbf x, t)$, respectively. Define
\begin{subequations}\label{eq:eBOMB}
\begin{equation} \label{eq:u}
    \mathbf u := (1 - \alpha) \mathbf v + \alpha \mathbf w.
\end{equation}
The trajectory $\mathbf x_i(t)$ of the $i$th clump of a raft obeys:
\begin{equation} \label{eq:eBOMBxdot}
    \dot{\mathbf x}_i = \mathbf u\vert_i + \tau \mathbf u_\tau\vert_i  +  \tau \mathbf F_i, 
\end{equation}
where
\begin{equation}
    \mathbf u_\tau := R\frac{D\mathbf v}{Dt} + R \left( f + \tfrac{1}{3}\omega \right)\mathbf v^\perp -  \frac{D\mathbf u}{Dt} - \left(f + \tfrac{1}{3}R \omega \right) \mathbf u^\perp,
\end{equation}
where
\begin{equation}
    \frac{D\mathbf v}{Dt} := \partial_t\mathbf v + (\mathbf v\cdot\nabla)\mathbf v
\end{equation}
and similarly for $\frac{D\mathbf u}{Dt}$. Here, $f = f_0 + \beta y$ is the Coriolis parameter (twice the local Earth's angular speed); $\perp$ means $\frac{\pi}{2}$-anticlockwise rotation; $\omega = -\nabla\cdot\mathbf v^\perp$ is the vertical component of the ocean velocity's vorticity; and $\mathbf F_i$ denotes an external force. Note that for a raft consisting of $N$ clumps, we obtain a system of $2N$ first-order ordinary differential equations, coupled by $\mathbf F_i$.

The force $\mathbf F_i$ felt by the clump labeled $i$ is given by
\begin{equation} \label{eq:F-spring-def}
    \mathbf F_i := - \sum_{j \in \neigh{i, t}} k(|\mathbf x_{ij}|) (|\mathbf x_{ij}| - L) \frac{\mathbf x_{ij}}{|\mathbf x_{ij}|},
\end{equation}
where $\mathbf x_{ij} := \mathbf x_i - \mathbf x_j$.  Here, $\neigh{i, t}$ is equal to the set of indices of clumps that are connected to $i$ at time $t$ and
\begin{equation} \label{eq:k}
    k(|\mathbf x_{ij}|) := \frac{A}{\mathrm{e}^{(|\mathbf x_{ij}| - 2L)/\Delta} + 1},
\end{equation}
\end{subequations}
where $\Delta$ is taken small enough such that $k(|\mathbf x_{ij}|) \approx A$ for $0\le |\mathbf x_{ij}| \le 2L$ and $k(|\mathbf x_{ij}|) \approx 0$ for $|\mathbf x_{ij}| > 2L$.  In this model, $\mathbf F_i$ serves as a restoring force that preserves the linkage between the clumps up to a certain distance, after which the clumps completely detach.

The equation governing the trajectory of a raft's center of mass, denoted $\mathbf x_\text{CM}(t) = \smash{\frac{1}{N}\sum_{i=1}^N}\mathbf x_i(t)$, remains unidentified, and this equation is what we aim to determine. To illustrate the difficulty inherent in the learning process, Eq.\@~\eqref{eq:eBOMBxdot} implies that
\begin{equation}
    \dot{\mathbf x}_\text{CM} = \sum_{i=1}^N \mathbf u\vert_i + \tau \mathbf u_\tau\vert_i,
\end{equation}
where we have used the fact that spring forces are equal and opposite among clumps. Note that this equation is formally independent of elastic interactions, but since it is not closed with respect to $\mathbf x_\text{CM}(t)$, it still depends on them. It follows that a model for $\mathbf x_\text{CM}(t)$ is not achievable in the most general case but an approximation may be obtained as long as the constituent clumps are sufficiently concentrated near the center of mass.

\section{Discovering the center of mass dynamics}\label{sec:dis}

The eBOMB model \eqref{eq:eBOMB} represents a nonautonomus $2N$-dimensional nonlinear dynamical system:
\begin{equation} 
    \dot{\mathbf x}_i = \mathbf f_i(\mathbf x_1,\mathbf x_2,\dotsc,\mathbf x_N,t),\quad i = 1,2,\dotsc,N,
    \label{eq:fi}
\end{equation}
where $\mathbf f_i$ is as given on the right-hand-side of \eqref{eq:eBOMBxdot}.  We assume that the trajectories of the elemental clumps forming a \emph{Sargassum} raft remain close to their center of mass trajectory, $\mathbf{x}_\text{CM}(t)$. Under these conditions, $\mathbf{x}_\text{CM}(t)$ is expected to approximately obey a nonautonomous two-dimensional nonlinear dynamical system:
\begin{equation} 
    \dot{\mathbf x}_\text{CM} = \mathbf f_\text{CM}(\mathbf x_\text{CM},t)
    \label{eq:fCM}
\end{equation}
for certain $\mathbf f_\text{CM}$. Equation \eqref{eq:fCM} can be viewed, under the stated conditions, as a \emph{reduced-order model} for the complete motion of a \emph{Sargassum} raft, governed by \eqref{eq:eBOMB}, which expresses as in \eqref{eq:fi}. Based on observations of $\mathbf x_\text{CM}(t)$, our aim is to determine the right-hand side of \eqref{eq:fCM} by employing two machine learning techniques, which are evaluated and compared. These techniques are, as anticipated above, LSTM RNNs \cite{Hochreiter-Schmidhuber-97} and SINDy \cite{Brunton-etal-16}. The latter method generates results in symbolic form, in contrast to the noninterpretable black-box approach of the former.

The datasets employed for training comprise $\textbf x_\text{CM}(t)$ derived from four distinct configurations of the eBOMB model, computationally implemented within the \href{https://julialang.org/}{Julia} package \href{https://github.com/70Gage70/Sargassum.jl}{\texttt{Sargassum.jl}}. In every scenario, the ocean velocity ($\mathbf v$) is depicted through a synthesis of geostrophic flow, deduced from multisatellite altimetry observations of sea-surface height \cite{LeTraon-etal-98}, and Ekman (wind dirven) drift, prompted by wind reanalysis \cite{Dee-etal-11}. This synthesis is calibrated to correspond with the velocities of drifters from the NOAA Global Drifter Program \cite{Lumpkin-Pazos-07}, which are drogued (anchored) at a depth of 15 meters. The resultant product has been substantiated through experiments involving the tracking of small objects in the Atlantic \cite{Olascoaga-etal-20, Miron-etal-20-GRL}. The wind velocity (measured at a 10-meter elevation) $\mathbf w$ is generated by the European Centre for Medium-Range Weather Forecasts (ECMWF) Reanalysis v5 (ERA5), the most recent ECMWF reanalysis available \cite{Hersbach-etal-20}.

We have developed four datasets for training, each with progressively increased complexity, as detailed here. First, we acknowledge that mesoscale ocean vortices (eddies), which are characterized by diameters of 100 km or more, are ubiquitous and play a significant role in transport \cite{Robinson-83}. Second, consistent with this understanding, these eddies, particularly quasigeostrophic eddies \cite{Pedlosky-87} with rotationally coherent material boundaries that consequently resist breakaway filamentation \cite{Haller-etal-16}, contain local finite-time forward attractors within their interiors for BOM particles linked by spring forces under calm wind conditions \cite{Beron-Miron-20, Bonner-etal-24}. Third, consistent with this rigorous result, anticyclonic eddies of this type within the Caribbean Sea have been noted to effectively transport \emph{Sargassum} and, when destabilized by thermal interactions with irregular topography, contribute to their dispersion along the Central American coastline \cite{Andrade-etal-20}. Taking all these considerations into account, the rafts have been strategically initialized near the periphery of 23 coherent Lagrangian eddies, which were extracted from the ocean velocity data $\mathbf v$ throughout 2017, employing geodesic eddy detection \cite{Haller-Beron-13, Haller-Beron-14} as implemented in the \href{https://julialang.org/}{Julia} package \href{https://github.com/CoherentStructures/CoherentStructures.jl}{\texttt{CoherentStructures.jl}}. The trajectories of the four datasets are explicitly generated by integrating the eBOMB model \eqref{eq:eBOMB} for a period of 1.5 months with the following configurations:
\begin{itemize}
    \item[\textbf{\textsf A}] The wind is switched off and $\neigh{i, t}$ represents all possible connections to clump $i$ at time $t$. 
    \item[\textbf{\textsf B}] As in dataset \textbf{\textsf A}, but with the wind turned on.
    \item[\textbf{\textsf C}] The wind is switched off and $\neigh{i, t}$ is taken be a set of nearest-neighbors to clump $i$ at time $t$.
    \item[\textbf{\textsf D}] As in dataset \textbf{\textsf C}, but with the wind turned on. 
\end{itemize}

It should be noted that when $\neigh{i, t}$ corresponds to the nearest neighbors of the clump $i$ at time $t$ (datasets \textbf{\textsf C} and \textbf{\textsf D}) results in a decreased level of connectivity between the raft members compared to when $\neigh{i, t}$ encompasses all available connections to the clump $i$ at time $t$ (datasets \textbf{\textsf A} and \textbf{\textsf B}). Consequently, one can anticipate a more extensive dispersion of clumps due to advection driven by ocean currents and winds in the former scenario than in the latter. The parameters used have been optimized for the observed \emph{Sargassum} distributions, and are listed in Table 2 of Bonner et al. \cite{Bonner-etal-24}. The only observation is that the natural length of the spring $L$ is calculated based on the initial clump setup to ensure that the model is properly scaled. In our simulations, each raft comprises 100 clumps, covering an area of 50 km by 50 km.

Finally, each of the two machine learning methodologies, LSTM RNNs and SINDy, involves a distinct implementation. In every case, one of the 23 trajectories produced is reserved for testing, while the remaining 22 are employed in training.

\subsection{The LSTM RNN approach}\label{sec:LSTM}

Our implementation of the LSTM RNN is inspired by the method developed by Wan and Sapsis \cite{Wan-Sapsis-18} and subsequently applied within an oceanographic context by Aksamit et al. \cite{Aksamit-etal-20}. 

The main idea involves considering a \emph{reference} nonautonomous vector field $\mathbf f_\text{ref}(\mathbf x,t)$ and attempting to learn corrections to it via an LSTM RNN. Upon examining the eBOMB model \eqref{eq:eBOMB}, it becomes evident that the most straightforward selection for $\mathbf f_\text{ref}$ is $\mathbf u$, which is the weighted average of ocean velocity and wind, as defined in \eqref{eq:u}. However, this approach proves impractical due to the unknown nature of the windage parameter $\alpha$ a priori. Therefore, we have chosen to consider $\mathbf f_\text{ref} = \mathbf v$, where $\mathbf v$, the ocean velocity, in this instance is represented by observational data as outlined in the preceding section. Consequently, the formal expression 
\begin{equation}
    \mathbf f_\text{CM}(\mathbf x_\text{CM},t) \approx \mathbf v(\mathbf x_\text{CM},t) + \mathbf G(\mathbf x_\text{CM}, t)
    \label{eq:fCM-lstm}
\end{equation} 
is constructed, wherein $\mathbf G$ represents the term to be learned. We designate $\dot{\mathbf x}_\text{CM} = \mathbf v(\mathbf x_\text{CM},t) + \mathbf G(\mathbf x_\text{CM}, t)$ as the \emph{LSTM model}. It is crucial to acknowledge that $\mathbf G$ will \emph{not} be obtained in an explicit form, aligning with the inherent lack of interpretability in machine learning techniques based on NNs, such as LSTM RNNs.  

An artificially NN would aim to minimize the \emph{loss} between input or \emph{predictor}, given in our case by $\dot{\mathbf x}_\text{CM}(t) - \mathbf v(\mathbf x_\text{CM}(t),t)$, and the output or \emph{target}, defined by $\mathbf G(\mathbf x_\text{CM}(t),t)$.  This can be expressed as
\begin{equation}
    \min_\theta\|\dot{\mathbf x}_\text{CM}(t) - \mathbf v(\mathbf x_\text{CM}(t),t) - \mathbf G(\mathbf x_\text{CM}(t),t;\theta)\|^2,
\end{equation}
where $\|\,\|$ is an appropriately chosen norm and $\theta$ parametrizes the weights and biases that define the specific architecture of the NN. These parameters are iteratively optimized using gradient descent combined with backpropagation, while activation functions introduce nonlinearity, allowing the NN to learn complex patterns. These concepts and others such as neurons and layers along with specific aspects of LSTM RNNs (for instance, how the vanishing gradient problem, which affects the ability of RNNs to learn long-term dependencies, is addressed through the introduction of gates, or how time delays are internally accounted for, enabling to reconstruct dynamics from partial observations, e.g., as discussed in Wan et al. \cite{Wan-etal-18}) are reviewed in Appendix A of the Supplemental Online Material.

The eBOMB model \eqref{eq:eBOMB} suggests that the correction term to be learned, $\mathbf{G}$, in the most general scenario when wind is present, specifically in datasets \textbf{\textsf{B}} and \textbf{\textsf{D}}, should represent some function of the collection of variables:
\begin{equation}
    \left\{\mathbf v, \mathbf w, \frac{D\mathbf v}{Dt}, \frac{D\mathbf w}{Dt}, \omega\right\}.
    \label{eq:lib-lstm}
\end{equation}
These variables are the only directly observable data available. In fact, the weighted average velocity $\mathbf{u}$ cannot be directly observed due to the lack of prior knowledge of the windage, as noted earlier. Similarly, the restorative forces linking the elemental clumps that form a raft remain unobservable.  A important observation is that the dependence on the set of variables \eqref{eq:lib-lstm}, which is not directly accessible, is inherently \emph{highly nonlinear} owing to the intrinsic structure of NNs. The data-driven function $\mathbf G$ can be viewed as a map from these variables, evaluated along $\mathbf x_\text{CM}(t)$, thereby representing a time series, to the time series $\dot{\mathbf x}_\text{CM}(t) - \mathbf v(\mathbf x_\text{CM}(t),t)$.

The LSTM RNN's predictor and target are constructed using the time series $\mathbf{x}_\text{CM}(t)$ derived from datasets \textbf{\textsf{B}} and \textbf{\textsf{D}}, or alternatively from datasets \textbf{\textsf{A}} and \textbf{\textsf{C}}. In the most general case where wind is included, corresponding to datasets \textbf{\textsf{B}} and \textbf{\textsf{D}}, the dimension of the predictor is 9. In contrast, when wind is excluded, that is, in datasets \textbf{\textsf{A}} and \textbf{\textsf{C}}. the predictor dimension decreases to 5, since $\mathbf{w} = 0$ and $\frac{D\mathbf{w}}{Dt} = 0$ in that case. The dimension of the target consistently equals 2 regardless of the scenario.  

In this paper, we examine and contrast both shallow and deep configurations of LSTM RNNs:   
\begin{itemize}
    \item The \emph{shallow} configurations consist of a sequence input layer followed by a single LSTM layer with increasing numbers of neurons, succeeded by a dropout layer designed to mitigate overfitting, and a regression layer that fully connects the input to the output. The selection of the number $n$ of neurons was deliberately aligned with powers of 2, namely, $n = 2^1=2, 2^2 = 4, \dotsc 2^{10} = 1024$.  This approach aligns with established methodologies; alternative selections did not yield results that were substantially different.  The dropout layer was set to 20\%.  
    
    \item The \emph{deep} configurations include more than one LSTM layer, each one followed by a 20\% dropout layer.  We initiated the study with an incremental approach, starting with a single layer containing $\{2^1 = 2\}$ neurons. Subsequently, layers were added containing $\{2^2 = 4, 2^1 = 2\}$ neurons, followed by layers with $\{2^3 = 8, 2^2 = 4, 2^1 = 2\}$ neurons, extending up to those with $\{2^{10} = 1024, 2^9 = 512, \dotsc, 2^1 = 2\}$ neurons. Comparable results were observed when varying the number of neurons in incremental steps of 2 and 5. However, the specific architectural configuration of the layers was maintained consistently, and no efforts were made to investigate alternative structural arrangements.
\end{itemize}

In each case, the suitable normalized input data was partitioned into training sets comprising 70\% of the data and validation sets consisting of the remaining 30\% of the data, following a standard procedure.  Adaptive moment estimation (known as Adam) was selected as the optimization algorithm for training the LSTM RNN. The loss given network prediction and target values was set to mean squared error (mse), i.e., computed based on the squared $L_2$ norm. A maximum of 1500 epochs was specified, with a minibatch size of 50, incorporating shuffling at every epoch and sequence padding in the left direction to preclude the network from predicting padding. The initial learning rate was established at 0.01. The validation frequency was configured to 50, accompanied by a patience parameter set to 3. Root mean squared error (rmse) was identified as the metric for evaluation. A custom early stopping mechanism was implemented to terminate training prematurely when the training loss fell below a target loss threshold, set at 0.01. Otherwise, the default parameters of the function \href{https://www.mathworks.com/help/deeplearning/ref/trainnet.html}{\texttt{trainnet.m}} from the \href{https://www.mathworks.com/products/deep-learning.html}{Matlab's Deep Learning Toolbox}, employed in our study, were followed.  The training process consistently converged after approximately 250 iterations.

\begin{figure}
    \centering
    \includegraphics[width=\linewidth]{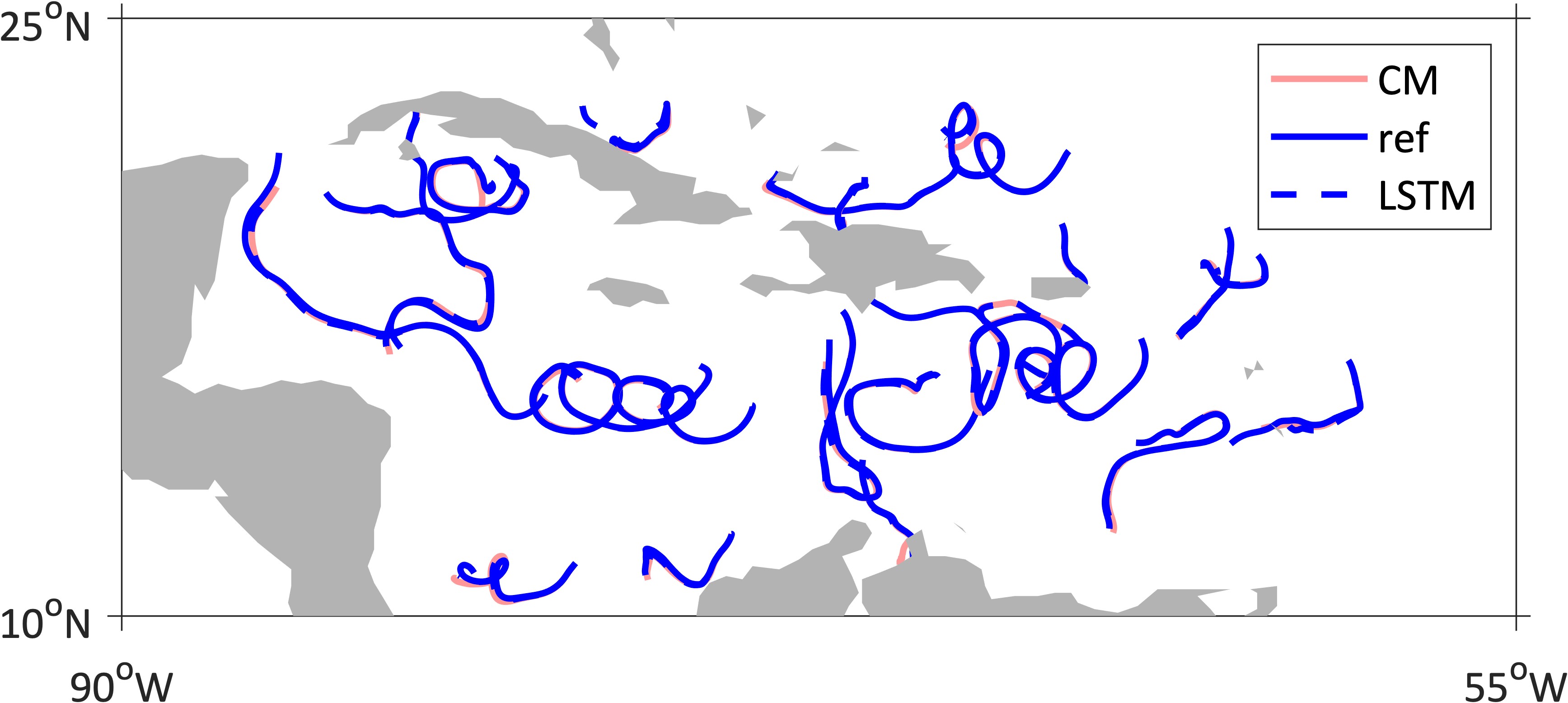}\\
    \includegraphics[width=\linewidth]{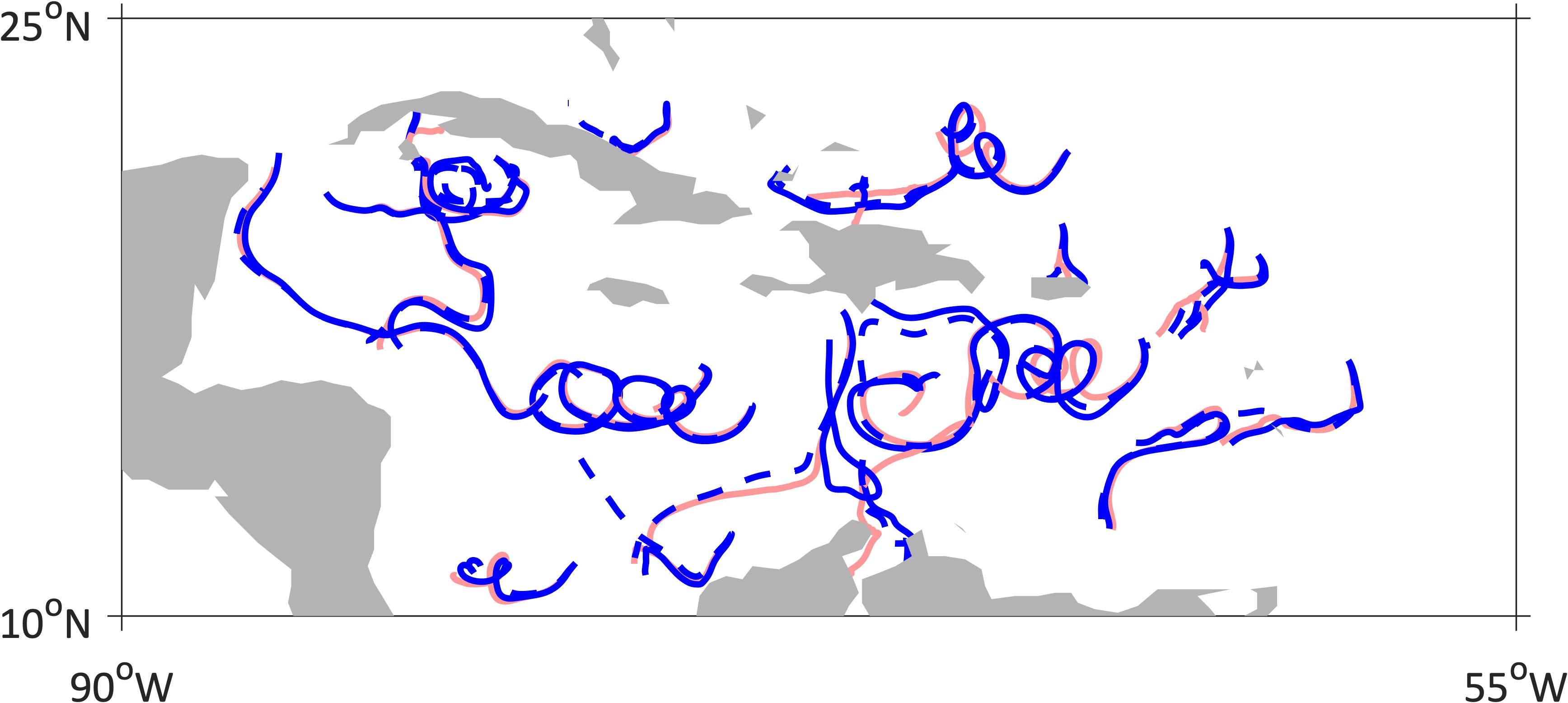}\\
    \includegraphics[width=\linewidth]{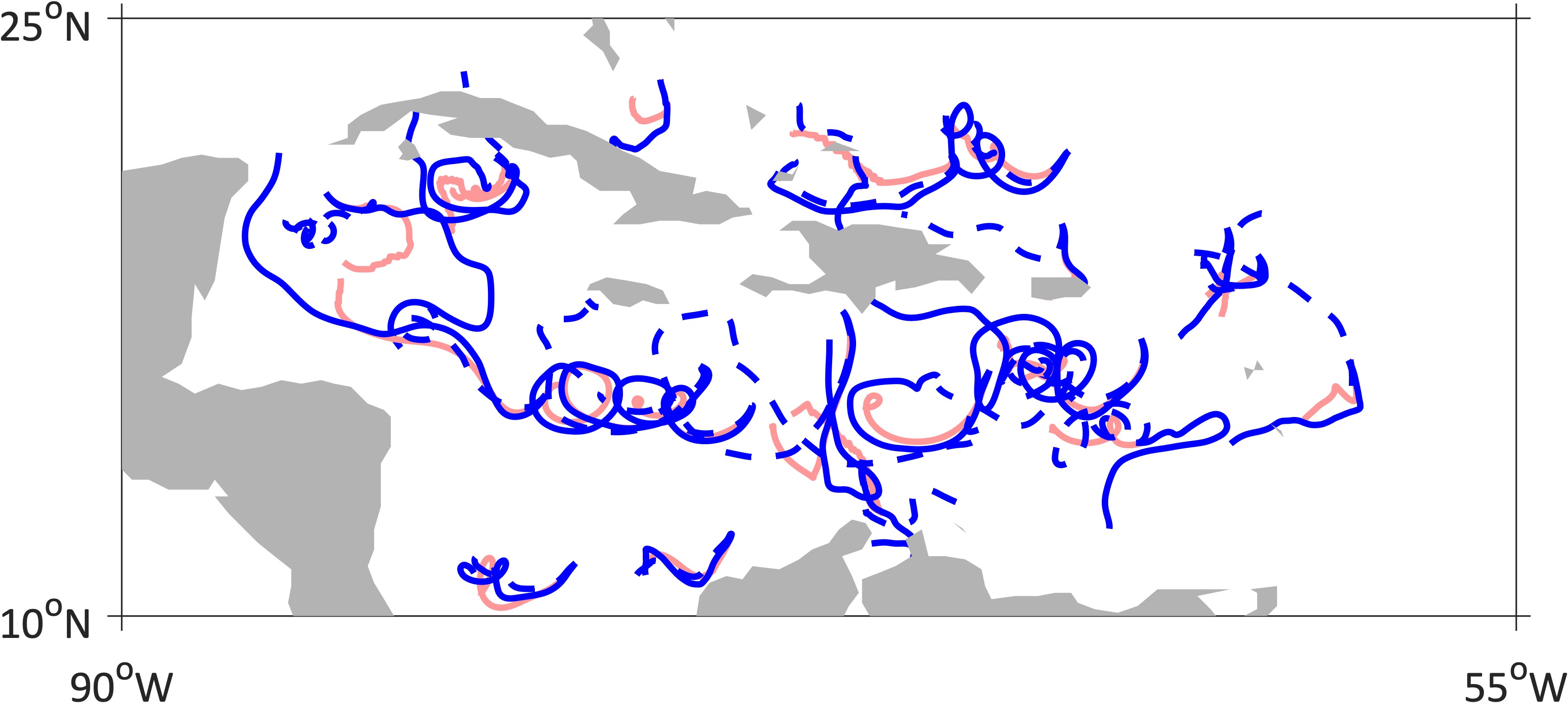}\\
    \includegraphics[width=\linewidth]{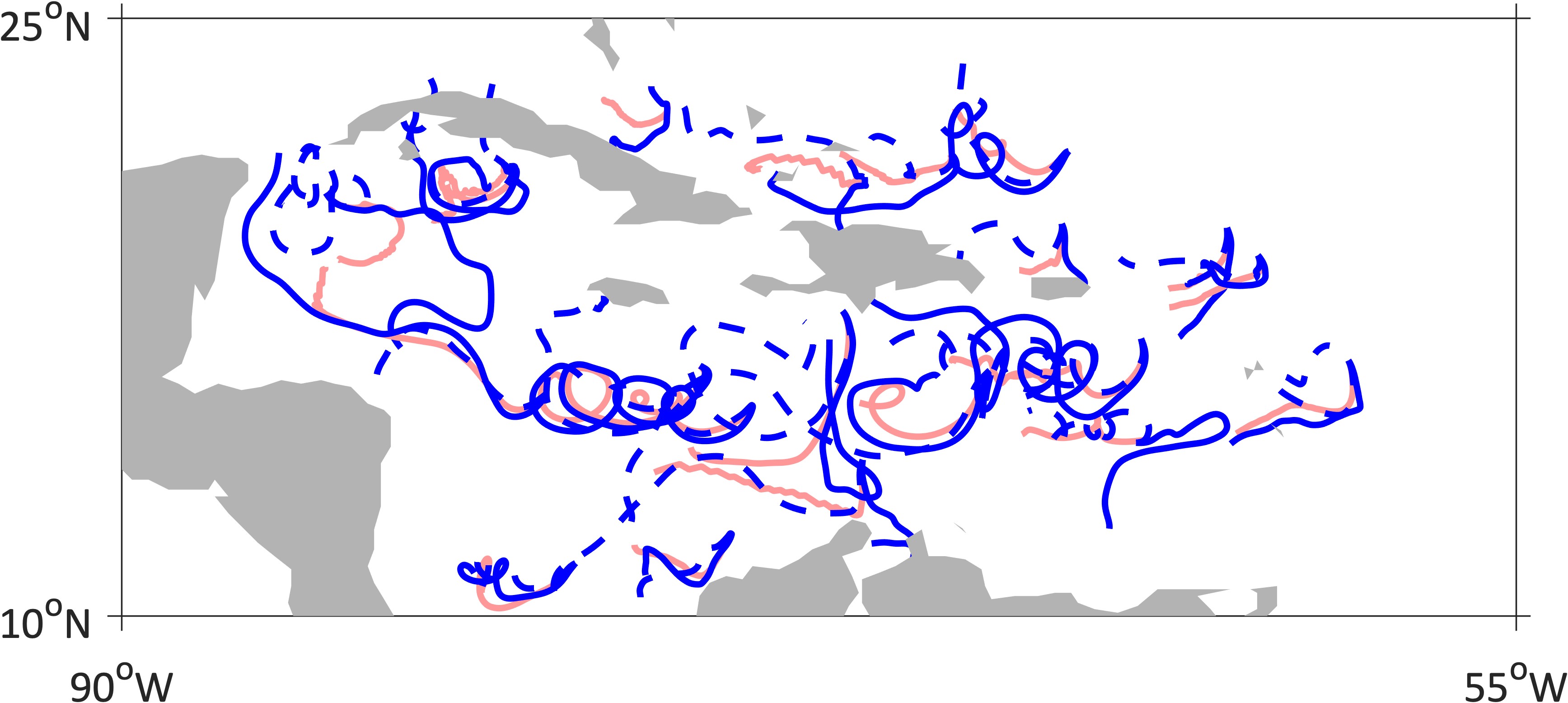}
    \caption{Center of mass trajectories (CM) of \emph{Sargassum} rafts evolving according to the eBOMB model, taken as true trajectories for LSTM RNN training, are shown beside reference trajectories (ref) and trajectories learned by the LSTM RNN (LSTM). From top to bottom, these trajectories correspond to datasets \textbf{\textsf{A}} through \textbf{\textsf{D}}. The LSTM RNN is configured to minimize the average Fr\'echet distance between the true and learned trajectories (cf.\ Fig.\@~ref{fig:dist}).}
    \label{fig:train}
\end{figure}

We start our exploration of the LSTM RNN methodology focusing on Fig.\@~\ref{fig:train}. This figure depicts the center of mass trajectories (CM), which represent the true trajectories employed during the training phase of the LSTM RNN, in conjunction with the reference trajectories (ref) and the trajectories learned by the LSTM RNN (LSTM). From top to bottom, the CM trajectories pertain to the datasets \textbf{\textsf{A}} through \textbf{\textsf{D}}. We recall that the duration of each trajectory is of 1.5 months. By reference trajectories, we refer to those resulting from the integration of $\dot{\mathbf x} = \mathbf v(\mathbf x,t)$ from initial conditions matching those of each CM trajectory, $\mathbf x_\text{CM}(t)$, at the corresponding temporal point, $t_0$, extending to $t_N$, the terminal time of $\mathbf x_\text{CM}(t)$.  LSTM trajectories are obtained by integrating the learned LSTM model, conceptually represented by $\dot{\mathbf x} = \mathbf v(\mathbf x,t) + \mathbf G(\mathbf x,t)$, employing identical initial conditions. This merits an explanation, which was previously omitted in earlier implementations \cite{Wan-Sapsis-18, Wan-etal-18, Aksamit-etal-20} of the LSTM RNN methodology, which served as inspiration for the current application. For each $\mathbf x_\text{CM}(t)$ incorporated into the training process, the learned LSTM RNN is used to construct the correction term sequence $\mathbf g(t_i) = \mathbf G(\mathbf x_\text{CM}(t_i),t_i)$, $i = 0, 1, \dotsc, N$. The Matlab function \href{https://www.mathworks.com/help/deeplearning/ref/dlnetwork.predict.html}{\texttt{dlnetwork.predict}}, employed here, performs this operation given the LSTM RNN learned by mapping the set of base variables \eqref{eq:lib-lstm} evaluated at $\mathbf x_\text{CM}(t_i)$ into $\mathbf g(t_i)$.  An interpolant is then generated, enabling the numerical integration of $\dot{\mathbf x} = \mathbf v(\mathbf x,t) + \mathbf g(t)$ from $\mathbf x(t_0) = \mathbf x_\text{CM}(t_0)$ over $t\in [t_0,t_N]$. 

\begin{figure}
    \centering
    \includegraphics[width=\linewidth]{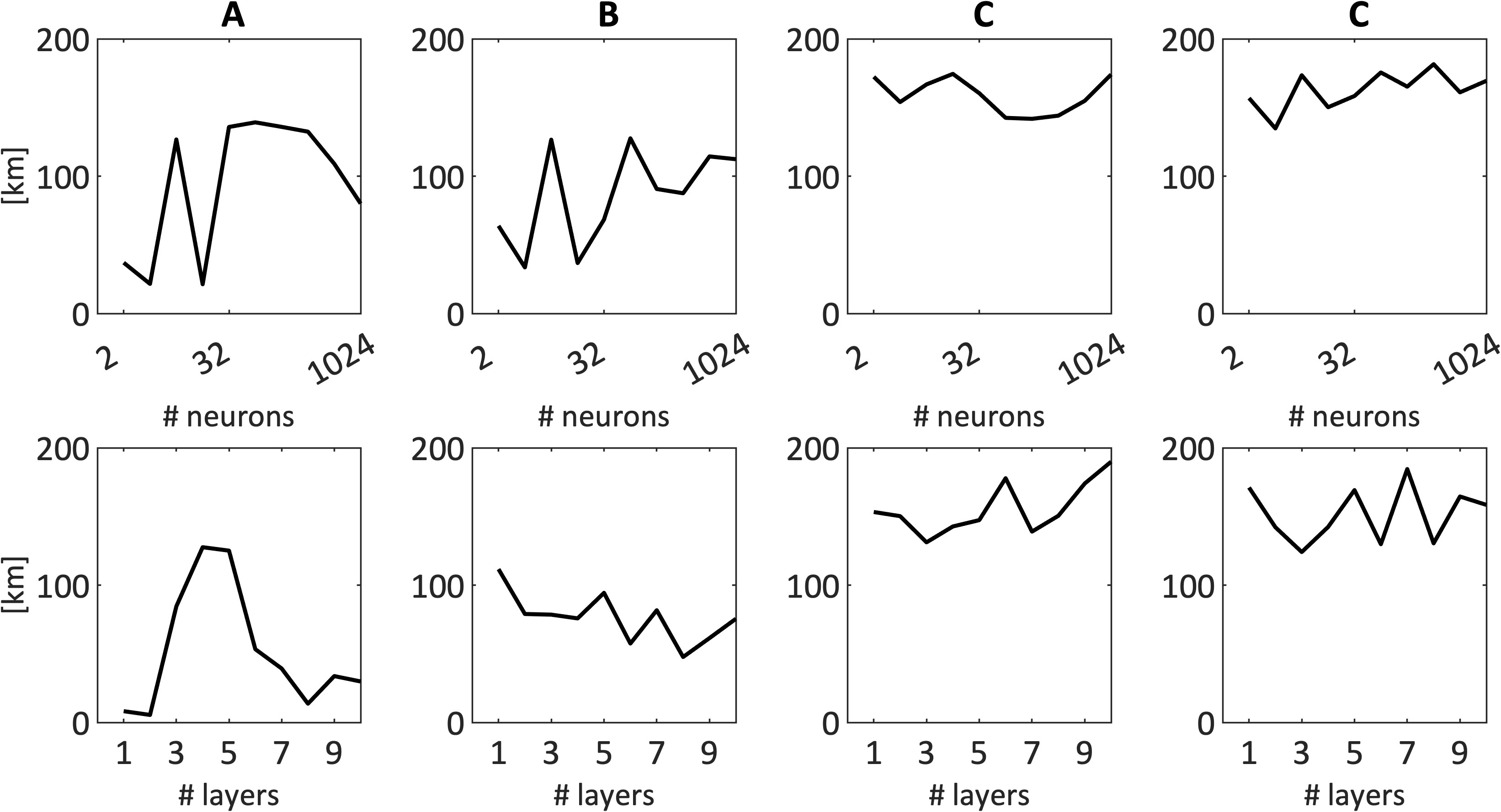}
    \caption{The average Fr\'echet distance between true and learned trajectories is shown for the different datasets, depending on the shallow LSTM RNN configuration (top panels) and the deep LSTM RNN configuration (bottom panels; cf.\ details in the text).}
    \label{fig:dist}
\end{figure}

\begin{figure*}
    \centering
    \includegraphics[width=\linewidth]{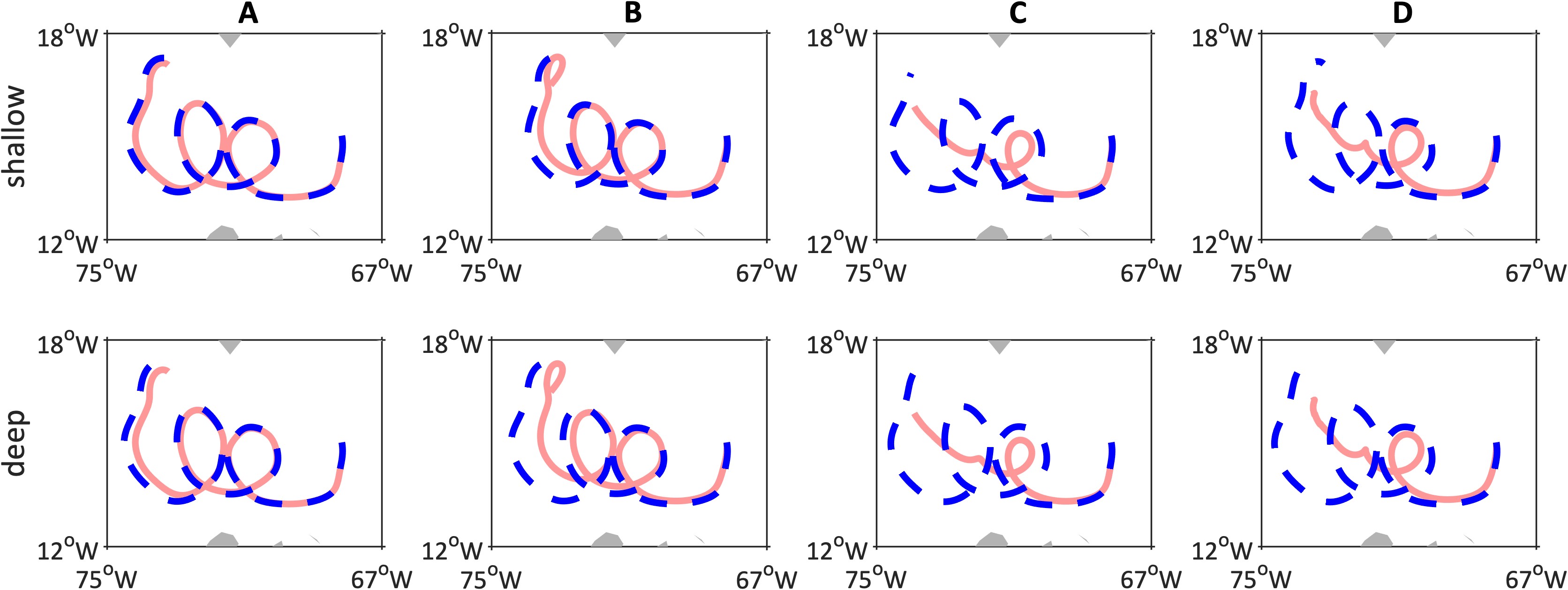}
    \caption{True trajectories (light red) not used during training and predicted trajectories (dashed blue) for the LSTM RNN architecture that minimizes the average Fr\'echet distance between true and learned trajectories during training (cf.\ Fig.\@~\ref{fig:dist}).}
    \label{fig:pred-lstm}
\end{figure*}

The selection of the LSTM RNN configuration for generating the learned trajectories depicted in each panel of Fig.\@~\ref{fig:train} was intentionally made to ensure the minimization of the average Fr\'echet distance, defined as the smallest of the maximum pairwise distances between points, between the true and learned trajectories. Figure \ref{fig:dist} illustrates this distance across various datasets, depending on the LSTM RNN configuration, with the shallow configuration presented in the upper panels and the deep configuration in the lower panels.  The initial observation to be made is that the Fr\'echet distance exhibits a consistent increase from \textbf{\textsf A} through \textbf{\textsf D}. This observation suggests that as the elementary clumps constituting the \emph{Sargassum} raft's begin to disperse from their center of mass---whether due to weakened inter-clump connections, the influence of wind, or a combination of these factors---it becomes increasingly challenging to accurately learn their dynamics.  An additional observation is that increasing the number of neurons within the shallow LSTM RNN configuration or increasing the number of layers in the deep configuration does not necessarily lead to improved learning performance.  This phenomenon is most apparent in the case of dataset \textbf{\textsf A}, which appears to be the most amenable to learning, as evidenced by the Fr\'echet distance transitioning from relatively minimal values to more substantial values, akin to those observed for datasets \textbf{\textsf C} and \textbf{\textsf D}, which are considered to be the most challenging to learn.

We finally proceed to evaluate the predictive capability of the trained LSTM model. This assessment focuses on the trajectory $\mathbf x_\text{CM}(t)$ from various center of mass trajectory datasets that was not used during the training phase. The predicted trajectories are generated by employing the trained LSTM RNN in a closed-loop mode, wherein preceding predictions are used as inputs for subsequent time steps in a sequence. Specifically, the construction of $\mathbf g(t_i)$ described above is conducted in a time-stepwise manner. Initially, starting from the base variables \eqref{eq:lib-lstm}, evaluated at $\mathbf x_\text{CM}(t_0)$, these are mapped into $\mathbf g(t_0)$ via the LSTM RNN using \href{https://www.mathworks.com/help/deeplearning/ref/dlnetwork.predict.html}{\texttt{dlnetwork.predict}}. Subsequently, $\dot{\mathbf x} = \mathbf v(\mathbf x, t) + \mathbf g(t_0)$ is integrated from $\mathbf x_\text{CM}(t_0)$ for a single time step to $t_1$ to yield $\mathbf x_1$. The base variables \eqref{eq:lib-lstm} evaluated at $\mathbf x_1$ are subjected to the same process iteratively until $t_N$, the terminal time of $\mathbf x_\text{CM}(t)$. In Fig.\@~\ref{fig:pred-lstm}, we depict the predicted trajectories, represented in dashed blue, evaluated in this manner, in comparison to the true trajectories $\mathbf x_\text{CM}(t)$ obtained from the various datasets constructed by integrating the eBOMB model \eqref{eq:eBOMB}, depicted in light red. In each scenario, the LSTM RNN architecture employed is the one that minimizes the average Fr\'echet distance between true and predicted trajectories during training. This observation further demonstrates that as the elementary clumps comprising the \emph{Sargassum} rafts begin to disperse from their center of mass, the task of accurately modeling the center-of-mass dynamics becomes progressively difficult.  

Although we have investigated the impact of various LSTM RNN architectures on learning the center-of-mass dynamics of \emph{Sargassum} rafts, our exploration is not comprehensive. Despite the limitations in scope, we have attempted to address this area. Such efforts have not been reported in previous LSTM RNN applications that inform our research. It remains possible that some architectures might more readily learn center-of-mass dynamics, particularly in scenarios where the constituent clumps of a raft diverge significantly from their center of mass. We ultimately conclude that it is feasible, with sufficient effort, to model such dynamics when the individual clumps are adequately interconnected, even when accounting for wind effects that potentially mitigate the anticipated attraction by mesoscale coherent Lagrangian eddies on \emph{Sargassum} rafts. However, the strength of interconnections between the \emph{Sargassum} clumps remains uncertain, which prevents us from asserting the generalizability of the results.

\subsection{The SINDy approach}\label{sec:SINDy}

The SINDy framework explicitly learns the functional form of the dynamical system in question, making it manifestly interpretable. Domain-specific knowledge allows for the inclusion of expected terms while excluding unhelpful ones, thereby enhancing the training process. The tradeoff is that SINDy is unable to learn terms that are not explicitly proposed, and hence may require explicit construction of bespoke functions for more complicated systems, such as in our case. 

We propose an approximate model for the right-hand side of Eq.\@~\eqref{eq:fCM} of the form
\begin{equation}
    \mathbf f_\text{CM}(\mathbf x_\text{CM},t) \approx \sum_i \xi_i \mathbf L_i(\mathbf x_\text{CM},t),
    \label{eq:fCM-sindy}
\end{equation}
where $\{\xi_i\}$ are coefficients and $\mathbf L_i$ is the $i$th member of a \emph{library} of candidate functions, denoted $\mathcal L$. The core objective of SINDy is to construct a library consisting of terms that are believed relevant and use sparsification techniques to keep as few as possible while maintaining accuracy. We will refer to $\dot{\mathbf x}_\mathrm{CM} = \sum_i \xi_i \mathbf L_i(\mathbf x_\text{CM},t)$ as the \emph{SINDy model}.

Referring to Eq.\@~\eqref{eq:eBOMB}, $\mathbf{v}$ and $\mathbf{w}$, their material derivatives and the vorticity are all logical inclusions in $\mathcal{L}$. In order to attempt to capture the fact that $\mathbf f_\text{CM}(\mathbf x_\text{CM},t)$ is nonlocal, we construct \emph{windowed} velocity interpolants  $\mathbf{v}_W(\textbf x, t; \delta)$ with $\delta > 0$ by
\begin{subequations} \label{eq:window-itps}
 \begin{align}
     \mathbf{v}_W (\textbf x, t; \delta) := \frac{\int_{B(\textbf x'; \delta)} \textbf{v}(\textbf x', t) \,d\textbf x'}{\int_{B(\textbf x; \delta)} \,d\textbf x'},
  \end{align}
  where
  \begin{align}
     B(\textbf x; \delta) := \big\{(x',y') : |x - x'| \leq \delta,\, |y - y'| \leq \delta\big\},
  \end{align}
\end{subequations}
and similarly for $\mathbf{w}_W(\textbf x, t; \delta)$. This allows $\mathbf f_\text{CM}$ to partially ``feel'' the effects of its immediate surroundings. Hence, our library is
\begin{equation}
    \mathcal{L} := \left\{\mathbf v, \mathbf w, \frac{D\mathbf v}{Dt}, \frac{D\mathbf w}{Dt}, \omega, \mathbf v_W, \mathbf w_W\right\}.
\end{equation}
The terms related to $\mathbf w$ are omitted when learning the equations corresponding to datasets $\textbf{\textsf A}$ and $\textbf{\textsf C}$. We also note that it is straightforward to include higher-order polynomial functions of constituents of $\mathcal{L}$, but we only include linear terms for ease of interpretability and speed of integration. Experimentation showed that these higher-order terms are significantly disfavored compared to the linear ones.  

The learning process proceeds in two main steps. First, Eq.\@~\eqref{eq:fCM-sindy} is converted into a regression problem,
\begin{equation}
    \mathbf{y} = \Theta \boldsymbol{\xi},
\end{equation}
where $\mathbf{y}$ is the \emph{target vector}, $\Theta$ is the \emph{feature matrix}, and $\boldsymbol{\xi}$ is the vector of \emph{coefficients} that needs to be determined. Second, the regression problem is solved with a sparsity-promoting algorithm. 

Given a trajectory $\mathbf x_\mathrm{CM}(t)$, sampled at discrete times $t_0, t_1, T-1$, a feature matrix $\Theta$ of size $T \times |\mathcal L|$ is constructed by evaluating $\mathcal{L}$ the discrete trajectory.  Two \emph{target creation algorithms} are then considered to construct the sparse regression problem, namely,
\begin{itemize}
    \item \emph{numerical differentiation}, in which case $\mathbf{y}_i = \dot{\mathbf x}_\mathrm{CM}(t_i)$, where the right-hand side is estimated using finite differences; and
    \item \emph{numerical integration} \cite{schaeffer2017sparse}, in which case $\mathbf{y}_i = \mathbf x_\mathrm{CM}(t_i) - \mathbf x_\mathrm{CM}(t_0)$ and $\Theta$ is replaced by $\smash{\int_{t_1}^{t_i}}\Theta\,dt$, where the integration, approximated using the the trapezoidal rule, is performed along rows of $\Theta$ (the coefficients in Eq.\@~\eqref{eq:fCM-sindy} are fixed).
\end{itemize}

We note that in the case where multiple disconnected trajectories are available, they can be separately converted to sparse regression problems and then concatenated into a single problem. To solve this, we consider \emph{sparse regression algorithms} from the following list.

\begin{itemize}
    \item \emph{Sequential Thresholding Least SQuares} (\emph{STLSQ}) \cite{brunton2016discovering}, where the feature matrix is iteratively regressed onto the target and $\boldsymbol\xi$ coefficients smaller than a prescribed threshold are removed.
    \item \emph{Automatic knee finding STLSQ} (\emph{AutoSTLSQ}), a refinement of the above where the threshold is selected automatically by finding the knee in a plot of cross-validation error versus threshold. 
    \item \emph{Least Absolute Shrinkage and Selection Operator} (\emph{LASSO}) \cite{tibshirani1996regression, chen2001atomic}, formulated as a convex quadratic programming problem by minimizing $\|\mathbf y - \Theta\boldsymbol\xi\|_2^2 + \lambda \|\boldsymbol\xi\|_1$ for a regularization parameter $\lambda$.
    \item \emph{Forward Stepwise regression} (\emph{ForwardStep}) \cite{efroymson1966stepwise}, where terms are iteratively added to the active set according to which provides the maximum correction of the current active set towards the target.
    \item \emph{Best subset selection} (\emph{Best}) \cite{Bertsimas2016}, which directly formulates the mixed integer optimization problem of minimizing $\|\mathbf y - \Theta\boldsymbol\xi\|_2^2$ subject to $\|\boldsymbol\xi\|_0 \leq k$ where the $\|\,\|_0$ denotes the $L_0$ (pseudo)norm. When applied to an arbitrary vector, it is equal to the number of nonzero elements in the vector. This algorithm is particularly advantageous as the final result will have at most $k$ terms for a prescribed $k$. Here, we use a fast, adaptive refinement of the basic algorithm \cite{zhu2020polynomial}.
\end{itemize}
Further information relating to the above algorithms is included in Appendix B of the Supplemental Online Material. 

Many other algorithms are available \cite{fasel2022ensemble, goyal2022discovery, wentz2023derivative}, but we do not consider them here as we are in a regime of relatively low noise and a small library. We learn a coefficient vector $\boldsymbol\xi$ for each pair of target creation and sparse regression algorithms. Given this vector, Eq.\@~\eqref{eq:fCM-sindy} is integrated and compared to the true trajectory. In order to evaluate the performance of each pair of algorithms, we apply the \emph{Special Information Criterion} (\emph{SIC}), defined by \cite{zhu2020polynomial}
\begin{equation} \label{eq:sindy-sic}
    \text{SIC}(\boldsymbol\xi) = n \log \mathcal F(\boldsymbol\xi) + \log(2) \|\boldsymbol\xi\|_0  \log \log n,
\end{equation}
where $n$ is the total number of training observations and $\mathcal F(\boldsymbol\xi)$ is the Fr\'echet distance between the integrated trajectory and the true trajectory. The SIC therefore rewards $\xi$ that are both sparse and accurate. 

\begin{figure*}
    \centering
    \includegraphics[width=\linewidth]{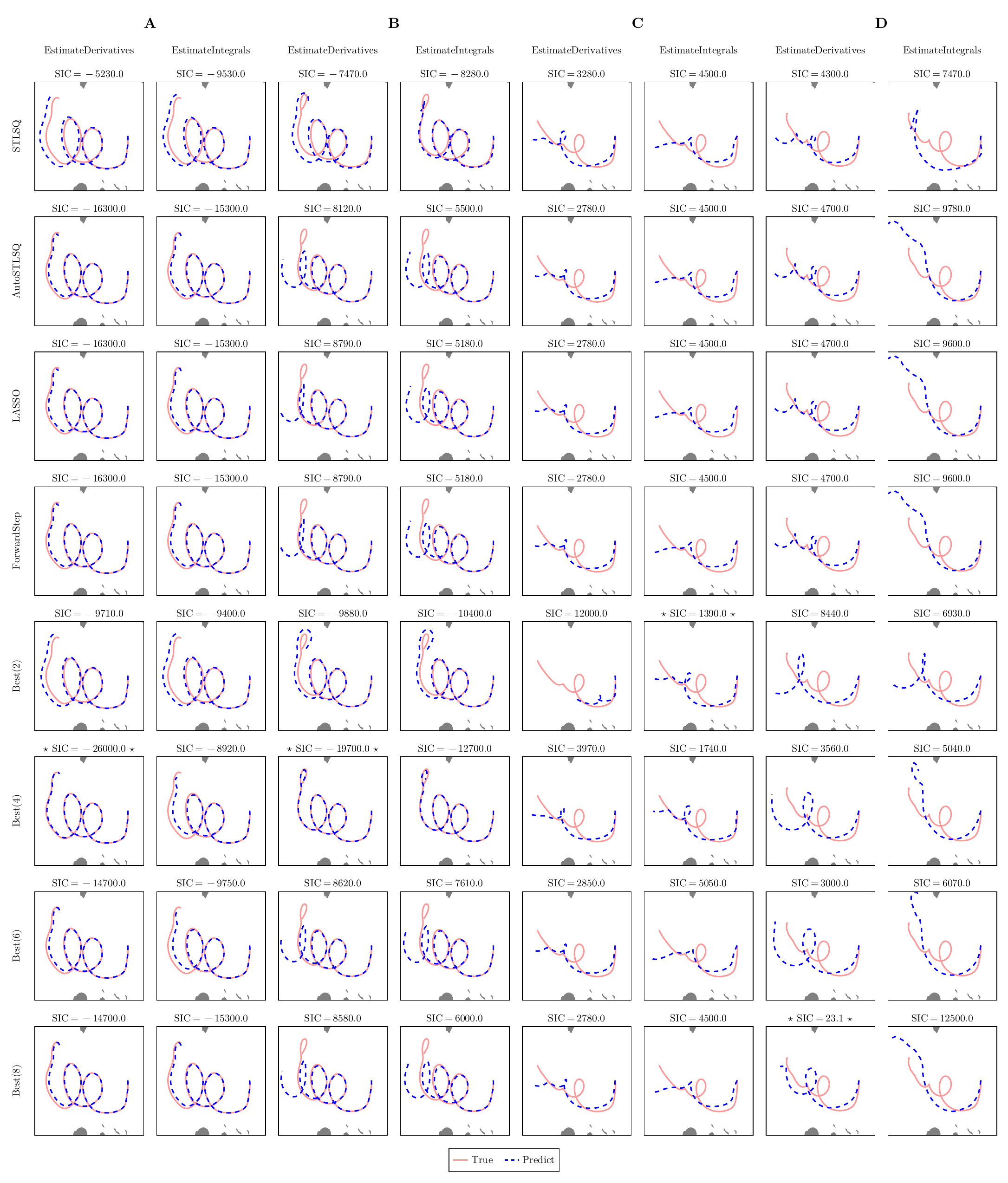}
    \caption{True (light red) and predicted (dashed blue) trajectories for each pair of target creation algorithms (columns) and sparse regression algorithms (rows) and each dataset \textbf{\textsf A}--\textbf{\textsf D} using the SINDy method. $\text{Best}(N)$ refers to the adaptive best subset selection algorithm with $N$ terms. Each panel shows the SIC computed using Eq.~\eqref{eq:sindy-sic}; the optimal value for each dataset is written as $\star \cdot \star$.}
    \label{fig:multialg-sindy}
\end{figure*}

Figure \ref{fig:multialg-sindy} presents the comparisons between the learned and true htrajectories. We apply the best subset selection algorithm for multiple subset sizes.

In general, no algorithm pair strictly outperforms the others but we select the $(\text{EstimateDerivatives}, \text{Best}(4))$ pair for our final estimates due to its strong performance averaged across all datasets. This pair returns the following SINDy models:
\begin{widetext}
\begin{subequations} \label{eq:sindy-learned-eqs}
\begin{align}
    \mathbf{f}_{\text{CM}}^{\text{\textbf{\textsf A}}} &= 
    \begin{pmatrix}
    0.924 & 0.00536 \\
    0 & 0.860
    \end{pmatrix} \mathbf{v} + 
    \begin{pmatrix}
    0.0769 & 0 \\
    -0.00516 & 0.155
    \end{pmatrix} \mathbf{v}_W +
    \begin{pmatrix}
    0& 0 \\
    0 & 0.0158
    \end{pmatrix} \frac{D \mathbf{v}}{D t}, \label{eq:fCM-eqs-A} \\
    \mathbf{f}_{\text{CM}}^{\text{\textbf{\textsf B}}} &= 
    \begin{pmatrix}
    0.901 & 0 \\
    0 & 0.856
    \end{pmatrix} \mathbf{v} + 
    \begin{pmatrix}
    0.00652 & 0 \\
    0 & 0.00298
    \end{pmatrix} \mathbf{w} + 
    \begin{pmatrix}
    0.105 & 0 \\
    0 & 0.154
    \end{pmatrix} \mathbf{v}_W +
    \begin{pmatrix}
    -0.00328 & 0 \\
    0 & 0
    \end{pmatrix} \mathbf{w}_W +
    \begin{pmatrix}
    0 & 0 \\
    0 & 0.0258
    \end{pmatrix} \frac{D \mathbf{v}}{D t}, \label{eq:fCM-eqs-B} \\
    \mathbf{f}_{\text{CM}}^{\text{\textbf{\textsf C}}} &= 
    \begin{pmatrix}
    -0.373 & 0 \\
    0 & -0.235
    \end{pmatrix} \mathbf{v} + 
    \begin{pmatrix}
    1.03 & 0.0906 \\
    -0.083 & 0.883
    \end{pmatrix} \mathbf{v}_W +
    \begin{pmatrix}
    0 & -0.142 \\
    0 & 0.184
    \end{pmatrix} \frac{D \mathbf{v}}{D t}, \label{eq:fCM-eqs-C} \\
    \mathbf{f}_{\text{CM}}^{\text{\textbf{\textsf D}}} &= 
    \begin{pmatrix}
    -0.303 & 0 \\
    0 & -0.289
    \end{pmatrix} \mathbf{v} + 
    \begin{pmatrix}
    -0.0213 & 0 \\
    -0.00218 & 0
    \end{pmatrix} \mathbf{w} + 
    \begin{pmatrix}
    0.927 & 0 \\
    0 & 0.889
    \end{pmatrix} \mathbf{v}_W +
    \begin{pmatrix}
    0.0296 & 0 \\
    0 & 0
    \end{pmatrix} \mathbf{w}_W +
    \begin{pmatrix}
    0 & 0.177 \\
    0 & 0
    \end{pmatrix} \frac{D \mathbf{v}}{D t}. \label{eq:fCM-eqs-D}
\end{align}
\end{subequations}
\end{widetext}

\begin{figure}
    \centering
    \includegraphics[width=\linewidth]{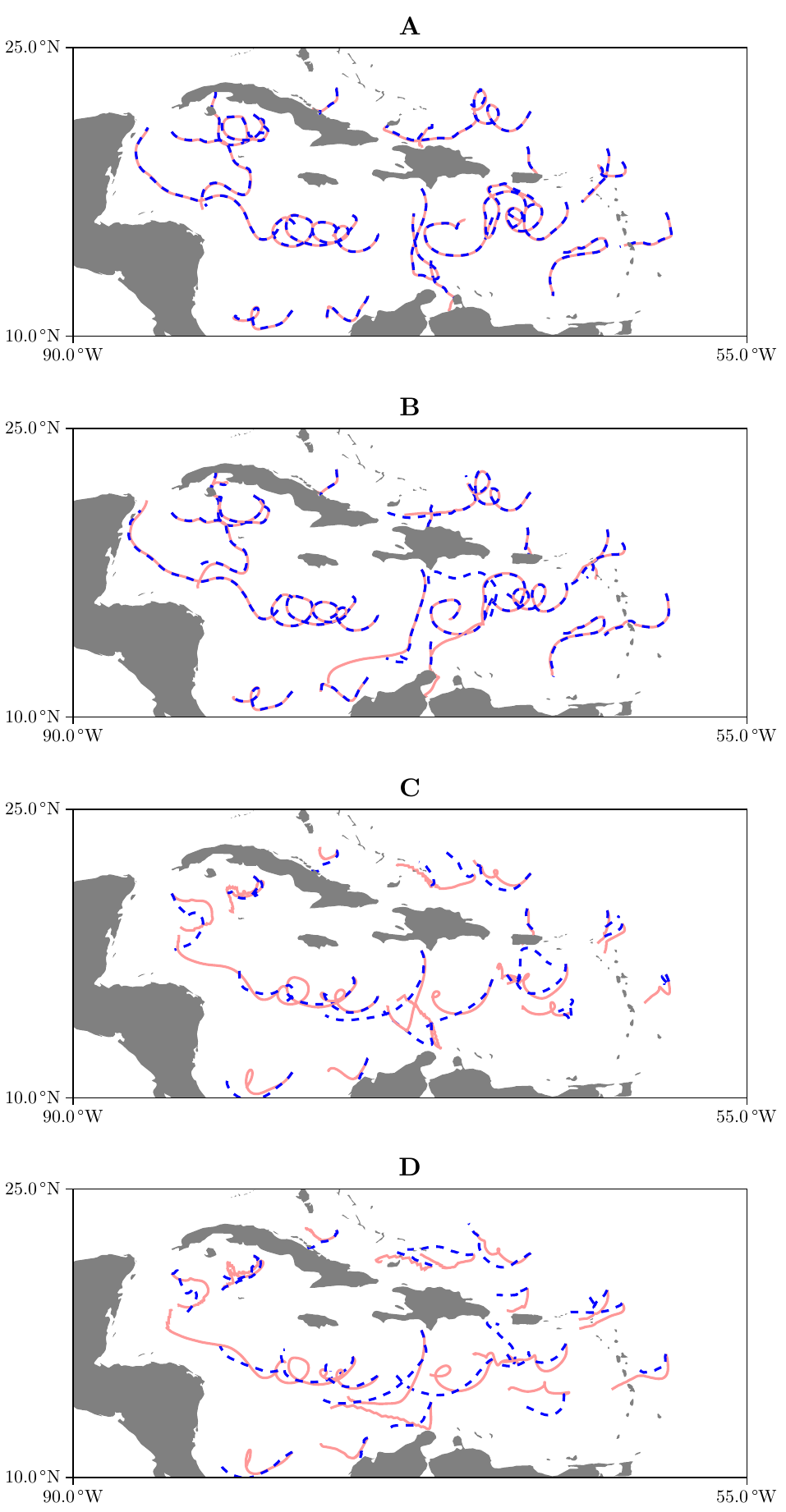}
    \caption{As Figure~\ref{fig:train}, but with learned trajectories obtained by integrating Eqs.~\eqref{eq:sindy-learned-eqs}.}
    \label{fig:train-sindy}
\end{figure}

Figure \ref{fig:train-sindy} shows trajectories obtained by integrating the learned equations compared to the reference trajectories used for training. We comment on the performance of the SINDy approach in each dataset, beginning with the fully connected, wind-free case \textbf{\textsf A}. In general, we expect that the more connected the raft is, and the closer each clump is on average to the center of mass, the more the trajectory of the center of mass will behave like the trajectory of a single particle. This is observed here, where nearly every algorithm has good performance and Eq.\@~\eqref{eq:fCM-eqs-A} is dominated by the diagonal matrix associated with $\mathbf{v}$. Small contributions from the nearly diagonal matrix associated with $\mathbf{v}_W$ provide the necessary corrections to account for the fact that even a very tightly clumped raft will not behave \emph{exactly} as a single particle. We also see a very small contribution from $\frac{D\mathbf v}{Dt}$, likely due to the fact that we report $\text{Best}(4)$. The SINDy algorithm could be tuned to increase its valuation of sparsity; we note, however, that the simpler algorithms tended to only find the $\mathbf{v}$ and $\mathbf{v}_W$ terms. In general, this demonstrates the success of the windowed velocity terms as we achieve high accuracy with no further library functions. 

Turning now the discussion to dataset \textbf{\textsf B}, we see the addition of a small diagonal matrix corresponding to $\mathbf{w}$. The effect of the windowed interpolants is much smaller in this case, and again we see a small asymmetric matrix associated with the material derivative. The contribution from wind in the true trajectories is small and difficult to resolve with the available data. We note that the predicted trajectory is nevertheless extremely accurate and, in fact, $\text{Best}(4)$ is the only algorithm that successfully captures the wind-induced loop at the end of the true trajectory.

We now turn to the discussion of dataset \textbf{\textsf C}, the more difficult case with nearest-neighbor connections. Examining the trajectories in Figure~\ref{fig:multialg-sindy}, we see that the performance across all algorithms is much worse than in datasets \textbf{\textsf A} and \textbf{\textsf B}. This is expected, as dataset \textbf{\textsf C} represents dynamics that are significantly more nonlocal. Nevertheless, $\text{Best}(4)$ captures the main features of the true trajectories. We see now that $\mathbf{v}_W$ is the dominant term, consistent with the requirement that an accurate model for the center of mass in this case must account for the increased spread of the individual clumps. In this case, we see a diagonal matrix associated with $\mathbf{v}$, where now its entries are negative. This suggests that the velocity itself acts as a correction to the windowed velocity, in particular, the specific form of $\mathbf{v}_W$ we use appears to overshoot the true trajectory if the correct features are captured. 

Finally, we consider dataset \textbf{\textsf D}, where wind has been activated. Yet again we see diagonal matrices associated with $\mathbf{v}$ and $\mathbf{v}_W$. In this case, the terms associated with wind appear to be mostly malformed---none enter into the meridional (``$y$'') component of the equations and the zonal (``$x$'') component have nearly equal and opposite coefficients. This is the most difficult of the four cases and although Eq.\@~\eqref{eq:fCM-eqs-D} produces a relatively accurate trajectory, it nevertheless lacks interpretability. The SINDy approach appears to be primarily limited by the quality of the library in this case. That is, superior target creation and sparse regression algorithms will likely not improve the results on datasets \textbf{\textsf C} and \textbf{\textsf D}.


\section{Discussion}\label{sec:disc}

The learned LSTM and SINDy models for the motion of the centers of mass of \emph{Sargassum} rafts perform generally similarly when applied to the four datasets considered. They give similar and quite accurate results in the cases of datasets \textbf{\textsf{A}} (tight clump connections, wind off) and \textbf{\textsf{B}} (tight clump connections, wind on), especially in the former case. Both models struggle equally when tested using datasets \textbf{\textsf{C}} (loose clump connections, wind of) and \textbf{\textsf{D}} (loose clump connections, wind on), especially in the latter case. This is consistent with the initial expectation that the centers of mass of the rafts can be controlled by a reduced-order model, expressed by the non-autonomous two-dimensional dynamical system \eqref{eq:fCM}, when the clumps that form a raft experience reduced dispersal, while keeping the raft sufficiently aggregated.

To a good approximation, the SINDy model for the case of dataset \textbf{\textsf{A}} is given by
\begin{equation}
    \dot{\mathbf x}_\text{CM} = \mathbf v + 0.1 \mathbf v_W.
\end{equation}
It is tempting to identify the learned SINDy model with the ``leeway'' model, which is commonly used in search-and-rescue operations at sea \cite{Breivik-etal-13}. However, $\mathbf{v}_W$ differs from the wind velocity, $\mathbf{w}$, as it represents a windowed ocean velocity. As noted above, this allows a raft's center of mass to ``sense'' the far-field flow rather than the immediate surrounding flow.  Consequently, it enables accounting for the effects among distant clumps, not just the neighboring ones. 

Multiple extensions to these approaches are possible. The SINDy approach in particular would benefit most from the creation of library functions that can handle greater dispersion. For example, one could consider time-dependent velocity windows that expand in order to model the increasing deviation of clumps from the center of mass. This technique is ultimately limited by the ability of a local function to capture nonlocal behavior. A promising direction would involve a hybrid approach which may use NNs to learn the appropriate combination of velocity interpolants to be included in a library which can then be sparsely regressed by SINDy. 

\section{Summary and conclusions}\label{sec:con}

In this paper, we have evaluated and compared two machine learning methodologies in terms of their ability to determine a governing equation for the motion of the center of mass of rafts of \emph{Sargassum} seaweed. These rafts move on the ocean surface under the influence of currents and winds while interacting elastically in a nonlinear manner. This motion is mathematically described by a Maxey--Riley model, referred to as the eBOMB model. However, the motion law for their centers of mass, which holds significant practical relevance, remains unresolved in a closed analytical form.

The methods considered are Long Short-Term Memory (LSTM) Recurrent Neural Networks (RNNs) and Sparse Identification of Nonlinear Dynamics (SINDy). LSTM RNNs are a type of neural network architecture designed to model sequential data and capture long-range dependencies. SINDy, on the other hand, is a method used for discovering governing symbolic equations from data under the assumption that the generically nonlinear dynamics admit a parsimonious (sparse) representation.  In each case, a closure modeling approach was taken, informed by the physics described by the eBOMB model.  

Both LSTM RNNs and SINDy demonstrated strong performance in scenarios with tightly connected clumps, the building blocks of a raft, and showed progressive degradation in accuracy with increasing complexity; that is, when windage is accounted for and less connected clumps are considered. The performance of the LSTM RNNs was highest when their architectures were kept simple, that is, when the number of neurons in shallow configurations and hidden layers in deep configurations were kept small. As is intrinsic to NNs, the LSTM RNNs provide black-box modeling capabilities that lack interpretability. SINDy achieved interpretability by explicitly learning functional forms using curated candidate function libraries. The incorporation of windowed velocity terms allowed us to successfully capture nonlocal interactions, particularly in datasets with less interconnected rafts.

Future work could explore integrating NNs with the SINDy method to enhance its capability in capturing nonlocal behaviors by adaptively expanding the library functions.

\section*{Acknowledgements}

The authors thank Mar\'ia J.\ Olascoaga (Rosenstiel School) for the benefit of many useful discussions about \emph{Sargassum} phenomenology.  FJBV expresses gratitude to Alan Kaptanglou (Courant Institute) for insightful discussions on data-driven modeling.

\section*{Funding}

This research was funded by the National Science Foundation (NSF) grant OCE2148499 and a grant from the Frost Institute for Data Science \& Computing (IDSC) of the University of Miami (UM) under the program ``Expanding the Use of Collaborative Data Science at UM.''

\section*{Author declarations}

\subsection*{Conflict of interest}

The authors have no conflict of interest to disclose.

\subsection*{Author contributions}

FJBV was responsible for the execution of the LSTM RNN calculations, while GB conducted the SINDy computations. The manuscript was collaboratively written by FJBV and GB.

\section*{Data availability}

The \href{https://julialang.org/}{Julia} package \href{https://github.com/70Gage70/Sargassum.jl}{Sargassum.jl} was developed by GB, while the \href{https://julialang.org/}{Julia} package \href{https://github.com/CoherentStructures/CoherentStructures.jl}{CoherentStructures.jl} was created by Daniel Karrasch. The synthesis of ocean velocities, derived from altimetry, wind, and drifter data, is obtainable through \href{ftp://ftp.aoml.noaa.gov/phod/pub/lumpkin/decomp}{ftp://ftp.aoml.noaa.gov/phod/pub/lumpkin/decomp}. The wind velocity data used originate from the ECMWF Reanalysis v5 (ERA5), which can be accessed via \href{https://www.ecmwf.int/en/forecasts/dataset/ecmwf-reanalysis-v5}{https://www.ecmwf.int/en/forecasts/dataset/ecmwf-reanalysis-v5}.

\appendix

\section{Maxey--Riley parameters of eBOM model}

According to the BOM equation, the windage ($\alpha$) varies with buoyancy ($\delta$) as
\begin{equation}
  \alpha(\delta) = \frac{\gamma\Psi(\delta)}{1 + (\gamma - 1)\Psi(\delta)}.
  \label{eq:alpha}
\end{equation}
Here, $\gamma$ is the air-to-water viscosity ratio, 
\begin{equation}
  \Psi(\delta) := \pi^{-1}\operatorname{acos}\Phi(\delta) - \pi^{-1}\Phi(\delta)\smash{\sqrt{1 - \Phi(\delta)^2}},
  \label{eq:Psi}
\end{equation}
giving the fraction of emerged particle's projected (in the flow direction) area,
\begin{equation}
  \Phi(\delta) :=
  \frac{1}{2}(\varphi(\delta)^{-1} + \varphi(\delta))
  +
  \frac{\mathrm{i}\sqrt{3}}{2}
  \left(\varphi(\delta) - \varphi(\delta)^{-1}\right),
\end{equation}
with the fraction of emerged particle piece's height given by $1 - \Phi(\delta)$, where
\begin{equation}
  \varphi(\delta) := \sqrt[3]{\mathrm{i}\sqrt{1 - (2\delta^{-1}
  - 1)^2} + 2\delta^{-1} - 1}.
\end{equation}
The Stokes time ($\tau$) is given by
\begin{equation}
    \tau(\delta,a) := \frac{1 - \frac{1}{6}\Phi(\delta)}{\big(1 + (\gamma - 1)\Psi(\delta)\big)\delta^4}\cdot \frac{a^2\rho}{\mu},
    \label{eq:tau}
\end{equation}
where $a$ is the particle radius, $\rho$ is the water density, and $\mu$ stands for viscosity. Lastly, parameter $R$ is given by
\begin{equation}
    R(\delta) := \frac{1 - \frac{1}{2}\Phi(\delta)}{1 - \frac{1}{6}\Phi(\delta)}.
    \label{eq:R}
\end{equation}

It is important to observe that Eqs.\@~\eqref{eq:alpha}, \eqref{eq:tau}, and \eqref{eq:R} apply within the range of $1 \le \delta \lessapprox 4$. For $\delta \ge 1$, Beron-Vera \cite{Beron-24-POFb} provides the corresponding expressions.


%

\end{document}